\documentclass[aps,showpacs,groupedaddress,superscriptaddress,twocolumn,longbibliography]{revtex4-1}

\usepackage{amsmath,amssymb}
\usepackage[utf8x]{inputenc}
\usepackage[english]{babel}
\usepackage{graphicx,epsfig}
\usepackage{todonotes}
\usepackage{hyperref}

\usepackage{braket}
\usepackage{soul}
\usepackage{color}
\makeatother
\usepackage{placeins}
\usepackage{todonotes}

\allowdisplaybreaks

\global\long\def\abs#1{\left|#1\right|}

\begin{document}
%
\title{Dynamical scaling of correlations generated\\ by short- and long-range dissipation}

\author{K. Seetharam}
\affiliation{Department of Electrical Engineering, Massachusetts Institute of Technologies, Cambridge, Massachusetts 02139, USA}
\affiliation{Department of Physics, Harvard University, Cambridge MA, 02138, USA}
\author{A. Lerose}
\affiliation{Department of Theoretical Physics, University of Geneva, 1211 Geneva, Switzerland}
\affiliation{SISSA, International School for Advanced Studies, via Bonomea 265, I-34136 Trieste, Italy}
\affiliation{INFN, Istituto Nazionale di Fisica Nucleare, Sezione di Trieste, I-34136 Trieste, Italy}
\author{R. Fazio}
\affiliation{International Center for Theoretical Physics ICTP, Strada Costiera 11, I-34151, Trieste, Italy}
\affiliation{Dipartimento di Fisica, Universita di Napoli 'Federico II', Monte S. Angelo, I-80126 Napoli, Italy}
\author{J. Marino}
\affiliation{Department of Physics, Harvard University, Cambridge MA, 02138, USA}
\affiliation{Institut f\"ur Physik, Johannes Gutenberg Universit\"at Mainz, D-55099 Mainz, Germany}

\begin{abstract}
We study the spatio-temporal spreading of  correlations in an ensemble of spins   due to dissipation characterized by short- and long-range spatial profiles. Such emission channels can be synthetized with tunable spatial profiles in lossy cavity QED experiments using a magnetic field gradient and a Raman drive with multiple sidebands. We consider systems initially in an uncorrelated state, and find that correlations widen and contract in a novel pattern intimately related to both the dissipative nature of the dynamical channel and its  spatial profile.  
Additionally, we make a methodological contribution by generalizing non-equilibrium  spin-wave theory to the case of dissipative systems and derive equations of motion for any translationally invariant spin chain whose dynamics can be described by a combination of Hamiltonian interactions and dissipative Lindblad channels.
Our work aims at extending the study of correlation dynamics to purely dissipative quantum simulators and compare them with the established paradigm of correlations spreading in Hamiltonian systems.  
\end{abstract}


\date{\today}
\maketitle

\section{Introduction}

%
%

%
%

%
%
%

A deep understanding of how correlations spread in quantum many-body systems can catalyze experimental developments and applications in quantum science and technology, ranging from quantum computation and simulation to quantum sensing.
In integrable closed many-body systems, correlations are paradigmatically understood to spread due to entangled pairs of quasiparticles in an initial non-equilibrium state: excitations travel at 
a finite velocity across the system, with quantum information thereby spreading in a linear light-cone~\cite{calabrese2006time,  mathey2010light, calabrese2011quantum, cheneau2012light, mitra2013correlation,  alba2017entanglement}. Such behavior is ubiquitous in generic short-range interacting systems~\cite{lieb1972finite} unless the propagation of quantum information is suppressed by slow dynamics or ergodicity breaking~\cite{AndersonLocalization,AbaninRMP,DeRoeck1,Schiulaz1,Cirac:QuasiMBLWithoutDisorder,SalaPRX20,SchultzStarkMBL,RefaelStarkMBL,SmithDisorderFreeLocalization,Kormos:2017aa,LeroseSuraceQuasilocalization}.  

%
%
Systems with long-range interactions circumvent the constraints imposed by locality and permit remote degrees of freedom to build up correlations which respect only a milder notion of causality~\cite{hastings2006spectral, HaukeTagliacozzo, kastner, EisertLRCorrelations, vodola2014kitaev, Gorshkov1, Gorshkov2, MatsutaLRLRbound,  Daley, lerose2020origin, ElseLRLRbound,TranPRX20Hierarchy,Tran21LRlightcone}.  Specificaly, in such systems, the effect of a local perturbation does not generally decay exponentially fast outside a linear light-cone.  This feature makes long-range interactions an important ingredient in several theoretical and experimental topics of current interest, such as fast quantum-state transfer~\cite{EldredgePRL17,TranPRX20Hierarchy} and fast scrambling dynamics~\cite{Guo20Signaling,BelyanskyPRL20Minimal}. Additionally, the cooperative nature of dynamics in long-range interacting systems earns them a special place in the realization of exotic nonequilibrium states of matter~\cite{LeroseKapitza,RussomannoDTCLMG,MachadoPRX20LongRangePrethermalPhases}.

Both short- and long-range interactions with variable strengths can be realized in several atomic and molecular platforms~\cite{blatt2012quantum,yan2013realizing, richerme2014non, zeiher2017coherent, moses2017new, kucsko2018critical,de2019observation}, as well as in optical platforms for simulating quantum many-body physics such as photonic waveguide, circuit QED, and cavity QED systems~\cite{black2003observation, majer2007coupling, leroux2010implementation, houck12,van2013photon,goban2014atom,thompson2013coupling,goban2015superradiance,eichler2015exploring,gonzalez2015subwavelength, marino2019spectrum, hung2016quantum,hosten2016quantum,MonikaPRL2019, Noh_2016,fitz,liu2017quantum,kollar2017supermode, leonard2017supersolid,landini2018formation,vaidya2018tunable, norcia2018cavity, noi, marino2019cavity, kroeze2018spinor,mivehvar2021cavity}.
Photonic or atomic losses are an essential aspect of these platforms, thus requiring coherent and dissipative dynamics to be treated on the same footing. 

The effect of local and collective dissipation on correlations spread by variable range coherent interactions have been addressed in a number of platforms at the interface of condensed matter and many-body quantum optics~\cite{Marino2012bis, PhysRevA.92.013603, PhysRevLett.120.020401, PhysRevLett.121.170402, Marinolong2012, PhysRevB.100.165144,PhysRevB.98.054302,paz2019critical,paz2021time}. Spatially extended dissipative processes, however, are more poorly understood although they can themselves generate correlations and have the potential to steer a quantum system into an entangled state just like coherent interactions~\cite{marino2021universality}. So far, studies of   dissipative dynamics have only focused on channels whose spatial profile has limited tunability~\cite{diehl,PhysRevA.86.013606, parmee2018phases, parmee2019decay}. 

Here, we explore how correlations spread due to  dissipation with a widely tunable spatial profile. Such a tunable dissipation channel exhibits novel spatio-temporal correlation patterns and can be implemented in cavity QED platforms~\cite{SeetharamShort}. In this work, we study a system of two-level atoms whose correlations are generated solely by a Markovian dissipation channel with a tunable spatial profile.  We consider both short- and long-range profiles with the goal of understanding whether quantum information propagates differently in such dissipative systems compared to their Hamiltonian counterparts, by a thorough analysis  of the spatio-temporal scaling built up by the former.

Spatially correlated emission naturally arises in atomic ensembles, where it manifests as cooperative phenomenon such as superradiance and subradiance~\cite{asenjo2017exponential,henriet2019critical}. These ensembles can be geometrically controlled to selectively emit into specified modes by tuning the mean atomic separation with respect to the photon wavelength~\cite{sierra2021dicke,masson2020many}. The tunability of correlated emission considered in our work, realized using a magnetic field gradient and a Raman drive with appropriately chosen sideband frequencies, can be considered a synthetic version of the geometric control in atomic ensembles. Cavity QED platforms with this synthetic control allow us to study the non-equilibrium dynamics of quantum correlations beyond conventional cooperative emission phenomenon.

We  consider spin systems which undergo semi-classical dynamics with quantum correlations either generated or destroyed by the dissipation channel, depending on the background collective motion of the spins. This dependence of the dissipative dynamics on the motion of the collective spin leads to a spatio-temporal correlation front which opens and then collapses. We are able to analyze the system in the thermodynamic limit by extending non-equilibrium spin-wave theory, previously developed for coherent Hamiltonian dynamics by two of the authors~\cite{LeroseShort,LeroseLong}, to the case of dissipative systems. This formalism has previously proved successful in treating a wide variety of nonequilibrium long-range interacting spin systems, allowing for the study of dynamical stabilization of exotic nonequilibrium ordered~\cite{LeroseKapitza} and time-crystalline~\cite{Zhu_2019,pizzi2021higher} phases, as well as the impact of quantum fluctuations on dynamical critical points~\cite{LeroseShort,LeroseLong}. 

The paper is organized as follows. In Sec.~\ref{sec:TDSW}, we present the formalism of nonequilibrium spin wave theory extended to dissipative systems, and derive equations of motion for any translationally-invariant spin chain undergoing a combination of coherent and dissipative dynamics when the dissipation can be described via Lindblad channels. This formalism constitutes the methodological core of our work. In Sec.~\ref{sec:model}, we introduce the specific spatially extended dissipation channel whose correlation dynamics we study in the remainder of the paper. The experimental implementation of this model with a tunable spatial profile is discussed in Ref.~\cite{SeetharamShort}. In Sec.~\ref{sec:DispDynamics}, we analyze the dynamical scaling of quantum correlations generated by this channel during transient non-stationary dynamics. In Sec.~\ref{fut}, we discuss future directions.


\section{Generalized nonequilibrium spin-wave theory}\label{sec:TDSW}

In this section, we derive the dissipative version of nonequilibrium spin-wave theory (NEQSWT).
This formalism allows us to obtain equations of motion for the relevant observables and their correlations in translationally-invariant spin chains governed by a master equation, such as the model, Eq.~\eqref{mainlind}, discussed in Sec.~\ref{sec:model}. 
Previously, NEQSWT has been used to study the non-equilibrium dynamics of a variety of unitary systems including interacting spin chains with competing short- and long-range interactions~\cite{ruckriegel2012time,LeroseShort,LeroseLong,pizzi2021higher}, variable-range interactions~\cite{LeroseKapitza,lerose2020origin,LeroseBridging}, and those coupled to a cavity mode~\cite{Zhu_2019}. Here, we extend the method to dissipative dynamics and derive equations of motion for any system whose dynamics is described by a combination of translationally-invariant Hamiltonians and translationally-invariant Lindblad channels. Our derivation can be used to construct equations of motion for the system described in Eq.~\eqref{mainlind}, and more generally for any translationally-invariant spin system whose dynamics is described by a master equation. 

The premise of NEQSWT is to assume that the system is well-described by a time-dependent strongly polarized collective spin, with a small number of spin-wave excitations on top of the collective polarization. The motion of the collective spin and the spin-waves are coupled, as the spin waves produce a back-reaction (or quantum feedback) that self-consistently modifies the mean-field trajectory of the collective spin. As the number of spin-waves is assumed to be small, we can treat the spins as bosons and the dynamics of the system is reduced to the motion of excitations on top of a moving ``condensate''. Formally, the treatment is a self-consistent time-dependent Hartree approximation of the lowest order Holstein-Primakoff expansion of the spin dynamics. The method is valid when the relevant excitations of the system are spin-waves and during the portion of dynamics in which the spin-wave population remains low. The advantage of NEQSWT is that it allows us to examine the dynamics of a thermodynamically large number of spins whenever the above two conditions are met. This typically results in control of dynamics over a time window significantly larger than what permissible with conventional low order Holstein-Primakoff expansions~\cite{altland2010condensed}.
Compared to straightforward~\cite{maghrebi2016nonequilibrium}, or cluster~\cite{jin2016cluster}, mean-field approaches, which can be unstable for driven-dissipative systems, NEQSWT can be considered a systematic improvement which enables the treatment of dissipative quantum many body dynamics using a method with a control parameter.


\subsection{Types of channels}

\label{sec_types}

We consider translationally-invariant spin systems described by a quantum master equation constructed from a combination of three types of channels,   each   characterized by a spin operator of the general form 
\begin{equation}
	\hat{L}_{n}=c_{x}^{F,U,D}\hat{S}_{n}^{x}+c_{y}^{F,U,D}\hat{S}_{n}^{y}+c_{z}^{F,U,D}\hat{S}_{n}^{z} \, .
	\label{eq_Ln}
\end{equation} 
The coefficients $ c_{x,y,z}^{F,U,D}$ take  arbitrary (complex) values, which can be chosen independently in the various channels of type $F$, $U$, and $D$ defined below. We assume $|c_x^{F,U,D}|^2+|c_y^{F,U,D}|^2+|c_z^{F,U,D}|^2=1$, so the magnitude of each channel is encoded in overall dimensionful coupling constants.

The first type of channel is unitary dynamics from a collective field  generated by the Hamiltonian
\begin{align}
\hat{H}_{F} & =\omega_{F}\sum_{n}\hat{L}_{n}.
\end{align}
Clearly, in order for $\hat H_F$ to be Hermitian, the coefficients $c^F_{x,y,z}$ appearing in the definition of operators $\hat L_n$ must be taken to be real.

The second type of channel is unitary dynamics with spatial character generated by a Hamiltonian
\begin{align}
\hat{H}_{U} & =\frac{\eta}{s\Gamma_{U,k=0}}\sum_{n,m}f_U\left(\left|n-m\right|\right)\left(\hat{L}_{m}^{\dagger}\hat{L}_{n}+h.c.\right)
\end{align}
where $\Gamma_{U,k}\equiv\sum_{r\in\left\{ -\frac{N}{2},\frac{N}{2}\right\} }e^{ikr}f_U\left(\left|r\right|\right)$ is the Fourier transform of the spatial profile $f_U\left(\left|n-m\right|\right)$, $N$ is the number of spins in the system, and $s$ is the total spin of each spin on the chain (typically taken to be $s=1/2$). The strength of this term is defined with a factor of $\Gamma_{U,k=0}$ as per the usual Kac renormalization that is used to normalize the contribution of this channel to dynamics in the case that $f_U\left(\left|n-m\right|\right)$ is long-range~\cite{kac}. 
The coefficients $c^U_{x,y,z}$ appearing in the definition of operators $\hat L_n$ may be complex in this case.
One can construct arbitrary unitary models of interest featuring two-body spin-spin interactions by combining various building-block Hamiltonians of the above forms, each defined through operators $\hat L_n$ of the form \eqref{eq_Ln} with different coefficients. For example, one can construct Heisenberg XYZ models with arbitrary spatial modulation of the couplings, including, as relevant limits,   one-axis and two-axis twisting Hamiltonians. 

The third type of channel is dissipative dynamics generated by a jump operator $\hat{L}_{n}$ of the form in Eq.~\eqref{eq_Ln}, with arbitrary complex coefficients $c^D_{x,y,z}$ chosen independently from those of the Hamiltonian channels. 
The contribution of this channel to an adjoint master equation for an operator $\hat{A}$ is
\begin{widetext}
\begin{align}
\mathcal{D}_{D}\left(\hat{A}\right) & =\frac{\kappa}{s\Gamma_{D,k=0}}\sum_{n,m}f_D\left(\left|n-m\right|\right)\left(\hat{L}_{n}^{\dagger}\hat{A}\hat{L}_{m}-\frac{1}{2}\left\{ \hat{L}_{m}^{\dagger}\hat{L}_{n},\hat{A}\right\} \right),
\end{align}
\end{widetext}
where we have once again renormalized the dissipative strength  with $\Gamma_{D,k=0}$ analogously to above. The usual cases of purely collective (i.e., fully permutationally invariant) dissipation can be recovered by choosing $f_D\left(\left|n-m\right|\right)=\delta_{n,m}$ for individual dissipation and $f_D\left(\left|n-m\right|\right)=\mathrm{constant}$ for collective dissipation. Note that the interaction matrix $f_D\left(\left|n-m\right|\right)$ for a valid Lindblad map must be positive semi-definite; this condition is violated if the same-site component of the spatial profile $f_D\left(\left|n-m\right|=0\right)$ vanishes. Therefore, a valid dissipative channel will always include a sufficiently strong local (diagonal) term. For this reason, the definition of couplings 
$f_U\left(\left|n-m\right|\right)=\abs{n-m}^{-\alpha}$
for $n\neq m$, usually taken for long-range Hamiltonian interactions, does not lead to a well-defined positive Lindblad generator.
In the following, we will thus include a hardcore parameter $R>0$ in the definition of our Lindblad generator spatial profile, entering as
\begin{equation}
    f_D(|r|) = \frac{1}{(R+|r|)^\alpha} \, .
\end{equation}
In Appendix~\ref{sec:appdx_positivity} we show that $R=1$ is sufficient to ensure positivity for all values of $\alpha$.

The dynamics of an operator $\hat{A}$  can then be expressed using an  adjoint master equation
\begin{equation}\label{eq:adjointMasterEq}
\frac{d}{dt}\hat{A}=\sum_{j}\frac{1}{i}[\hat{A},\hat{H}_{j}]+\sum_{j'}\mathcal{D}_{j'}\left(\hat{A}\right),
\end{equation}
where the sums run over Hamiltonians and dissipators of the types described above, 
F,each defined with different coefficients $c^{U,D}_{x,y,z}$.
As the system is translationally-invariant, we assume periodic boundary conditions and   define the Fourier transform of the spin components as $\hat{S}^{\alpha}_{k}=\sum_{n}e^{-ikn}\hat{S}^{\alpha}_{n}$ with $\alpha\in\{x,y,z\}$. The inverse transform is given by $\hat{S}^{\alpha}_{n}=\frac{1}{N}\sum_{k}e^{ikn}\hat{S}^{\alpha}_{k}$. The spins in Fourier space satisfy the commutation relation $[\hat{S}^{\alpha}_{k},\hat{S}^{\beta}_{k'}]=i\epsilon^{\alpha \beta \gamma}\hat{S}^{\gamma}_{k+k'}$.

We now rotate to a time-dependent frame defined by angles $\theta(t)$ and $\phi(t)$. Specifically, we apply the unitary transformation $\hat{V}\left(\theta,\phi\right)=e^{-i\phi\sum_{n}S_{n}^{z}}e^{-i\theta\sum_{n}S_{n}^{y}}$. Letting ${e_{\alpha}}$ be the unit vectors of the lab frame, the unit vectors of the rotated frame, ${e_{\tilde{\alpha}}}$, are given as
\begin{widetext}
\begin{align}
e_{\tilde{x}} =\left(\begin{array}{c}
\cos\theta\cos\phi\\
\cos\theta\sin\phi\\
-\sin\theta
\end{array}\right), \quad
e_{\tilde{y}} =\left(\begin{array}{c}
-\sin\phi\\
\cos\phi\\
0
\end{array}\right), \quad
e_{\tilde{z}} =\left(\begin{array}{c}
\sin\theta\cos\phi\\
\sin\theta\sin\phi\\
\cos\theta
\end{array}\right).
\end{align}
We will later choose $\theta(t)$ and $\phi(t)$ so that the z-axis of the rotated frame, $e_{\tilde{z}}$, aligns with the z-component of the collective spin $\hat{S}^{\tilde{\alpha}}=\sum_{n}\hat{S}^{\tilde{\alpha}}_{n}=\hat{S}^{\tilde{\alpha}}_{k=0}$. The cost of this time-dependent rotation is an additional `inertial' Hamiltonian
\begin{align}
\hat{H}_{\text{RF}} & =\sin\theta\dot{\phi}\hat{S}^{\tilde{x}}-\dot{\theta}\hat{S}^{\tilde{y}}-\cos\theta\dot{\phi}\hat{S}^{\tilde{z}}
\end{align}
that contributes to the dynamics. The three types of dynamical channels that   contribute to the dynamics of an operator $\hat{\tilde{A}}$ in the rotated frame take thus the form
\begin{align}
\hat{H}_{F}&=\omega_{F}\sum_{\tilde{\alpha}\in\left\{ \tilde{x},\tilde{y},\tilde{z}\right\} }F_{\tilde{\alpha}}\hat{S}_{k=0}^{\tilde{\alpha}}\label{eq:generatorUniformHam}\\
\hat{H}_{U}&=\frac{2\eta}{\Gamma_{U,k=0}Ns}\sum_{k}\Gamma_{U,k}\sum_{\tilde{\alpha},\tilde{\beta}\in\left\{ \tilde{x},\tilde{y},\tilde{z}\right\} }M^{U}_{\tilde{\alpha},\tilde{\beta}}\hat{S}_{-k}^{\tilde{\alpha}}\hat{S}_{k}^{\tilde{\beta}}\label{eq:generatorSpatialHam}\\
\mathcal{D}_{D}\left(\hat{\tilde{A}}\right)&=\frac{\kappa}{\Gamma_{D,k=0}Ns}\sum_{k}\Gamma_{D,k}\sum_{\tilde{\alpha},\tilde{\beta}\in\left\{ \tilde{x},\tilde{y},\tilde{z}\right\} }M^D_{\tilde{\alpha},\tilde{\beta}}\left(\hat{S}_{k}^{\tilde{\alpha}}\hat{\tilde{A}}\hat{S}_{-k}^{\tilde{\beta}}-\frac{1}{2}\left\{ \hat{S}_{-k}^{\tilde{\alpha}}\hat{S}_{k}^{\tilde{\beta}},\hat{\tilde{A}}\right\} \right)\label{eq:generatorSpatialDisp}
\end{align}
where we have defined
\begin{align}
F_{\tilde{\alpha}}\left(\theta,\phi\right) & =c^F_{x}G_{\tilde{\alpha},x}+c^F_{y}G_{\tilde{\alpha},y}+c^F_{z}G_{\tilde{\alpha},z}\\
M^{U,D}_{\tilde{\alpha},\tilde{\beta}}\left(\theta,\phi\right) & =\left((c^{U,D}_{x})^{*}G_{\tilde{\alpha},x}+(c^{U,D}_{y})^{*}G_{\tilde{\alpha},y}+(c^{U,D}_{z})^{*}G_{\tilde{\alpha},z}\right)\left(c^{U,D}_{x}G_{\tilde{\beta},x}+c^{U,D}_{y}G_{\tilde{\beta},y}+c^{U,D}_{z}G_{\tilde{\beta},z}\right)
\end{align}
\end{widetext}
and $G_{\tilde{\alpha}\beta}=e_{\tilde{\alpha}}\cdot e_{\beta}$ is the projection of the rotated basis vectors on the lab frame basis vectors. The choice of operators $\hat{L}_{n}$ are encoded in the coefficients $F_{\tilde{\alpha}}\left(\theta,\phi\right)$ or $M^{U,D}_{\tilde{\alpha},\tilde{\beta}}\left(\theta,\phi\right)$ while the choice of spatial profiles $f_{U,D}\left(\left|n-m\right|\right)$ are encoded in $\Gamma_{U,k}$, $\Gamma_{D,k}$. Note that the dynamics of the above channels does not decompose into independent dynamics for each wave vector $k$ as sectors of different momenta are coupled via the self-consistent feedback of the $k=0$ mode. 

\subsection{Holstein-Primakoff expansion in a moving vacuum}

We now bosonize the spins via a lowest-order Holstein-Primakoff transformation~\cite{altland2010condensed}
\begin{align}
\hat{S}_{n}^{\tilde{z}} =s-\hat{b}_{n}^{\dagger}\hat{b}_{n}, \quad
\hat{\tilde{S}}_{n}^{+} =\left(2s\right)^{\frac{1}{2}}\hat{b}_{n}, \quad
\hat{\tilde{S}}_{n}^{-} =\left(2s\right)^{\frac{1}{2}}\hat{b}_{n}^{\dagger}
\end{align}
where $\hat{b}_{n}^{\dagger}$ and $\hat{b}_{n}$ are bosonic creation and annihilation operators representing spin flips along the chain and satisfy canonical commutation relations
$\left[\hat{b}_{n},\hat{b}_{m}^{\dagger}\right]=\delta_{nm}$. In Fourier space, the mapping becomes
\begin{align}
\hat{S}_{k}^{\tilde{x}} &=\left(\frac{Ns}{2}\right)^{\frac{1}{2}}\left\{ \hat{b}_{k}+\hat{b}_{k}^{\dagger}\right\}, \\
\hat{S}_{k}^{\tilde{y}} &=\frac{1}{i}\left(\frac{Ns}{2}\right)^{\frac{1}{2}}\left\{ \hat{b}_{k}-\hat{b}_{k}^{\dagger}\right\}, \\
\hat{S}_{k}^{\tilde{z}} &=Ns\delta_{k,0}-\sum_{k'}\hat{b}_{k'}^{\dagger}\hat{b}_{k+k'}
\end{align}
where $\hat{b}_{k}^{\dagger}=\frac{1}{\sqrt{N}}\sum_{n}e^{ikn}\hat{b}_{n}^{\dagger}$ and $\hat{b}_{k}=\frac{1}{\sqrt{N}}\sum_{n}e^{-ikn}\hat{b}_{n}$ are bosonic creation and annihilation operators representing spin-wave excitations. It is useful to work in terms of quadrature operators $\hat{q}_{k}$ and $\hat{p}_{k}$ which are expressed in terms of the creation and annihilation operators as $\hat{b}_{k}^{\dagger}=\frac{1}{\sqrt{2}}\left(\hat{q}_{k}-i\hat{p}_{k}\right)$ and $\hat{b}_{k}=\frac{1}{\sqrt{2}}\left(\hat{q}_{k}+i\hat{p}_{k}\right)$. Note that these momentum space quadrature operators satisfy the commutation relation $\left[\hat{q}_{k},\hat{p}_{k'}\right]=i\delta_{k',-k}$. The mapping between spins and bosonic modes can be given in terms of the quadrature operators as
\begin{align}
\hat{S}_{k}^{\tilde{x}} &=\left(Ns\right)^{\frac{1}{2}}\hat{q}_{k}, \\
\hat{S}_{k}^{\tilde{y}} &=\left(Ns\right)^{\frac{1}{2}}\hat{p}_{k}, \\
\hat{S}_{k}^{\tilde{z}}  &=Ns\delta_{k,0}-\frac{1}{2}\sum_{k'}\left(\hat{q}_{k'}\hat{q}_{k-k'}+\hat{p}_{k'}\hat{p}_{k-k'}-\delta_{k,0}\right).
\end{align}
It is also useful to define
\begin{equation}
n_{k}=\langle\hat{b}_{k}^{\dagger}\hat{b}_{k}\rangle=\frac{1}{2}\langle\left(\hat{q}_{k}\hat{q}_{-k}+\hat{p}_{k}\hat{p}_{-k}-1\right)\rangle
\end{equation}
with $n_{k=0}$ corresponding to the condensate density and $n_{k\neq0}$ corresponding to the occupation of the spin-wave mode at wavevector $k$. The evolution of the $k=0$ mode represents the dynamics of the spin-wave vacuum and the evolution of the $k\neq0$ represents dynamics of spin-waves on top of the moving vacuum. In the thermodynamic limit, we can  treat the spin-wave vacuum classically~\cite{LeroseBridging,lerose2020origin}, while treating the spin-waves as quantum mechanical excitations. In practice, this amounts to replacing $\hat{S}^{\tilde{z}}_{k=0}$ by a c-number $\langle \hat{S}^{\tilde{z}}_{k=0}\rangle$ and using the total spin-wave density
\begin{multline}
\epsilon\left(t\right)=\frac{1}{Ns}\sum_{k\neq0} n_{k}\left(t\right) \\
=\frac{1}{Ns}\sum_{k\neq0} \frac{\langle\hat{q}_{k}\left(t\right)\hat{q}_{-k}\left(t\right)+\hat{p}_{k}\left(t\right)\hat{p}_{-k}\left(t\right)-1\rangle}{2}
\end{multline}
as a control parameter for the approximation.  The `time-dependent' part of NEQSWT references choosing the rotating frame angles $\theta(t)$ and $\phi(t)$ at every momentum in time so that the $\tilde{z}$ axis aligns with the collective spin, which amounts to determining the equations of motion for these angles by enforcing $\langle S^{\tilde{x}}_{k=0}\rangle=0$ and $\langle S^{\tilde{y}}_{k=0}\rangle=0$. The position of the collective spin on the Bloch sphere defined in the lab frame is given as $\vec{m}=(m^{x},m^{y},m^{z})$ where
\begin{align}
m^{x}(t)  &=\sin\theta(t)\cos\phi(t), \\
m^{y}(t) &=\sin\theta(t)\sin\phi(t), \\ 
m^{z}(t)  &=\cos\theta(t).
\end{align}
This choice extends the validity of spin-wave theory to larger window of dynamics by redefining the spin-wave vacuum, represented by the collective spin, at every point in time so that the total spin-wave density on top of the vacuum remains small~\cite{LeroseLong}. In the dilute regime of $\varepsilon(t)\ll1$, spin waves behave as free bosonic modes which scatter self-consistently only with the collective magnetization ($k=0$ mode).

As long as $\epsilon\left(t\right)$ remains small, the majority of angular momentum in the system resides in the collective $k=0$ mode (taken to be aligned with the $\tilde{z}$ axis) and higher order terms in the Holstein-Primakoff transformation can be ignored~\cite{LeroseLong,LeroseShort}. The system's dynamics can then be described as that of the collective spin on a Bloch sphere with a small density of spin-waves, negligibly reducing the length of this collective magnetization. TDSW is valid up to times $\sim 1/\epsilon^2$ (see for instance Refs.~\cite{LeroseLong,LeroseShort}). As a practical rule of thumb, the dynamics of spins are faithfully captured as long as the spin-wave density satisfies $\epsilon(t)\lesssim  0.2$ for the effects illustrated in Section~\ref{sec:DispDynamics}.

We apply the Holstein-Primakoff transformation described above to the adjoint master equation Eq.~\eqref{eq:adjointMasterEq}. A sufficiently small spin-wave density allows us to truncate the equations of motion for the system at the Gaussian level;  expectation values of operators that are more than quadratic in spin-wave operators are negligible in this limit. This approximation then allows for a closed set of non-linear coupled dynamical equations involving only the angles $\theta(t)$ and $\phi(t)$, representing the one-point correlation functions, and the two-point correlation functions defined below:
\begin{align}\label{eq:twopointFunctions}
\Delta_{k}^{qq}\left(t\right)   &=\left\langle \hat{q}_{k}\left(t\right)\hat{q}_{-k}\left(t\right)\right\rangle, \\
 \Delta_{k}^{pp}\left(t\right)   &=\left\langle \hat{p}_{k}\left(t\right)\hat{p}_{-k}\left(t\right)\right\rangle, \\
\Delta_{k}^{qp}\left(t\right)   &=\frac{1}{2}\left\langle \hat{q}_{k}\hat{p}_{-k}+\hat{p}_{k}\hat{q}_{-k}\right\rangle.
\end{align}
The dynamics of these two-point functions act as feedback for the motion of ${\theta}(t)$ and ${\phi}(t)$.

Specifically, we substitute the spin operators with bosonic creation and annihilation operators in the Hamiltonian or dissipator and keep contributions that are at most quadratic in bosonic operators. We then substitute quadrature operators for the creation and annihilation operators before computing equations of motion for $\hat{q}_{k=0}$, $\hat{p}_{k=0}$, $\hat{q}_{k}\hat{q}_{-k}$, $\hat{p}_{k}\hat{p}_{-k}$, and $\frac{1}{2}(\hat{q}_{k}\hat{p}_{-k}+\hat{p}_{k}\hat{q}_{-k})$. The first two quantities and enforcement of $\langle S^{\tilde{x}}_{k=0}\rangle=\langle S^{\tilde{y}}_{k=0}\rangle=0$ yields equations of motion for the angles $\theta(t)$ and $\phi(t)$ respectively, while the latter three quantities yield equations of motion for the two-point functions given in Eq.~\eqref{eq:twopointFunctions}. 

It is important to note three technical points. First, we must do the Gaussian approximation in terms of bosonic creation and annihilation operators rather than quadratures as $\hat{b}_{k}^{\dagger}\hat{b}_{k}$ is the quantity that is related to the small parameter $\varepsilon$ that we are expanding around; doing the approximation in terms of quadrature operators may yield spurious terms in the final equations due to zero-point quantum fluctuations. Second, we must apply the Holstein-Primakoff transformation and Gaussian approximation at the level of the generators Eqs.~\eqref{eq:generatorUniformHam}-\eqref{eq:generatorSpatialDisp} before calculating the equation of motion for an operator $\hat{\tilde{A}}$; performing the Gaussian approximation after computing the equation of motion may also introduce spurious terms in the final equations. Third, the chain rule for derivatives does not apply to operators evolving under a Lindblad master equation so the equations for the two-point functions must be directly computed~\cite{gardiner2004quantum}; we cannot construct these equations from a product of the equations of motion for the one-point functions as is commonly done when NEQSWT is applied to purely unitary systems.

\subsection{Equations of motion}

The content of this section is intended as a user guide for assembling equations of motion for arbitrary quantum master equations of the general form considered in this work, corresponding to adjoint master equations that can be expressed as Eq.~\eqref{eq:adjointMasterEq}. The following are a set of mechanical rules to construct the right-hand side of the equations of motion.

First, we start with the contributions of the Larmor Hamiltonian $\hat{H}_{\text{RF}}$ which will always be present due to the rotation of the reference frame:
\begin{equation} \label{eq:gaussiane}
\begin{split}
	\frac{d}{dt}\theta &=0\\
	\frac{d}{dt}\phi &=0\\
	\frac{d}{dt}\Delta_{k}^{qq} & =\cos\theta\dot{\phi}\left(2\Delta_{k}^{qp}\right)\\
	\frac{d}{dt}\Delta_{k}^{pp} & =-\cos\theta\dot{\phi}\left(2\Delta_{k}^{qp}\right)\\
	\frac{d}{dt}\Delta_{k}^{qp} & =-\cos\theta\dot{\phi}\left(\Delta_{k}^{qq}-\Delta_{k}^{pp}\right)
\end{split}
\end{equation}
Each channel $j$, given by a choice of one of the generators in Eqs.~\eqref{eq:generatorUniformHam}-\eqref{eq:generatorSpatialDisp}, then contributes to the above equations as
\begin{align}
\frac{d}{dt}\theta &\rightarrow \frac{d}{dt}\theta + d\theta_{j}\\
\frac{d}{dt}\phi &\rightarrow \frac{d}{dt}\phi + d\phi_{j}\\
\frac{d}{dt}\Delta_{k}^{qq} &\rightarrow \frac{d}{dt}\Delta_{k}^{qq} + dQ_{j}\\
\frac{d}{dt}\Delta_{k}^{pp} &\rightarrow \frac{d}{dt}\Delta_{k}^{pp} + dP_{j}\\\\
\frac{d}{dt}\Delta_{k}^{qp} &\rightarrow \frac{d}{dt}\Delta_{k}^{qp} + dW_{j}
\end{align}
Below we give the contributions to the equations of motion from each type of channel. It is useful to define the quantities
\begin{equation}
	\xi_{\tilde{\alpha},\tilde{\beta}}=\frac{M_{\tilde{\beta},\tilde{\alpha}}}{M_{\tilde{\alpha},\tilde{\beta}}}=\frac{M_{\tilde{\alpha},\tilde{\beta}}^{*}}{M_{\tilde{\alpha},\tilde{\beta}}}
\end{equation}
\begin{equation}
	\delta^{\eta\xi}=\frac{1}{\Gamma_{k=0}Ns}\sum_{k\neq0}\Gamma_{k}\Delta_{k}^{\eta\xi}.
\end{equation}
defined analogously for each superscript $U$ or $D$.

The contributions from a $\hat{H}_{F}$ channel are
\begin{align}
d\theta_{H_{F}} & =\omega_{F}F_{\tilde{y}}\\
d\phi_{H_{F}} & =-\omega_{F}F_{\tilde{x}}\frac{1}{\sin\theta}\\
dQ_{H_{F}} & =-2\omega_{F}F_{\tilde{z}}\Delta_{k}^{qp}\\
dP_{H_{F}} & =2\omega_{F}F_{\tilde{z}}\Delta_{k}^{qp}\\
dW_{H_{F}} & =\omega_{F}F_{\tilde{z}}\left(\Delta_{k}^{qq}-\Delta_{k}^{pp}\right)
\end{align}
To ease the notation, we drop the superscripts $U$ and $D$ in the coefficients in the following equations for the channels $\hat H_L$ and ${\mathcal{D}}_L$.
The contributions from a $\hat{H}_{U}$ channel are
\begin{widetext}
\begin{align}
d\theta_{H_{L}}  & =-M_{\tilde{x},\tilde{z}}4\eta\frac{1}{\Gamma_{k=0}Ns}\sum_{k'}\Gamma_{k'}\frac{1}{2}\left\langle \hat{q}_{-k'}\hat{p}_{k'}+\xi_{\tilde{x},\tilde{z}}\hat{p}_{-k'}\hat{q}_{k'}\right\rangle \nonumber \\
& +M_{\tilde{y},\tilde{z}}2\eta\left(1+\xi_{\tilde{y},\tilde{z}}\right)\left(1-\varepsilon-\delta_{\alpha}^{pp}-\frac{1}{Ns}n_{k=0}-\frac{1}{Ns}\Delta_{k=0}^{pp}\right)
\end{align}
\begin{align}
d\phi_{H_{L}}  & =M_{\tilde{y},\tilde{z}}\frac{1}{\sin \theta}4\eta\frac{1}{\Gamma_{k=0}Ns}\sum_{k'}\Gamma_{k'}\frac{1}{2}\left\langle \hat{p}_{-k'}\hat{q}_{k'}+\xi_{\tilde{y},\tilde{z}}\hat{q}_{-k'}\hat{p}_{k'}\right\rangle \\
& -M_{\tilde{x},\tilde{z}}\frac{1}{\sin \theta}2\eta\left(1+\xi_{\tilde{x},\tilde{z}}\right)\left(1-\varepsilon-\delta_{\alpha}^{qq}-\frac{1}{Ns}n_{k=0}-\frac{1}{Ns}\Delta_{k=0}^{qq}\right)
\end{align}
\begin{align}
dQ_{H_{L}}  & =M_{\tilde{y},\tilde{y}}\eta\cdot8\frac{\Gamma_{k}}{\Gamma_{k=0}}\Delta_{k}^{qp}-M_{\tilde{z},\tilde{z}}\eta\cdot8\Delta_{k}^{qp}+M_{\tilde{x},\tilde{y}}4\eta\left(1+\xi_{\tilde{x},\tilde{y}}\right)\frac{\Gamma_{k}}{\Gamma_{k=0}}\Delta_{k}^{qq}
\end{align}
\begin{align}
dP_{H_{L}}  & =-M_{\tilde{x},\tilde{x}}\eta\cdot8\frac{\Gamma_{k}}{\Gamma_{k=0}}\Delta_{k}^{qp}+M_{\tilde{z},\tilde{z}}\eta\cdot8\Delta_{k}^{qp}-M_{\tilde{x},\tilde{y}}4\eta\left(1+\xi_{\tilde{x},\tilde{y}}\right)\frac{\Gamma_{k}}{\Gamma_{k=0}}\Delta_{k}^{pp}
\end{align}
\begin{align}
dW_{H_{L}}  & =-M_{\tilde{x},\tilde{x}}\eta\cdot4\frac{\Gamma_{k}}{\Gamma_{k=0}}\Delta_{k}^{qq}+M_{\tilde{y},\tilde{y}}\eta\cdot4\frac{\Gamma_{k}}{\Gamma_{k=0}}\Delta_{k}^{pp}\\
& +M_{\tilde{z},\tilde{z}}\eta\cdot4\left(\Delta_{k}^{qq}-\Delta_{k}^{pp}\right)
\end{align}
The contributions from a $\mathcal{D}_{D}$ channel are
\begin{align}
d\theta_{\mathcal{D}_{D}} & =-iM_{\tilde{x},\tilde{z}}\frac{1}{2}\kappa\frac{1}{\Gamma_{k=0}Ns}\sum_{k'}\Gamma_{k'}\left\langle \hat{q}_{-k'}\hat{p}_{k'}-\xi_{\tilde{x},\tilde{z}}\hat{p}_{k'}\hat{q}_{-k'}\right\rangle \\
& -iM_{\tilde{y},\tilde{z}}\frac{1}{2}\kappa\left(1-\xi_{\tilde{y},\tilde{z}}\right)\left(1-\varepsilon+\delta_{\alpha}^{pp}-\frac{1}{Ns}n_{k=0}+\frac{1}{Ns}\Delta_{k=0}^{pp}\right)
\end{align}
\begin{align}
d\phi_{\mathcal{D}_{D}} & =iM_{\tilde{y},\tilde{z}}\frac{1}{\sin \theta}\frac{1}{2}\kappa\frac{1}{\Gamma_{k=0}Ns}\sum_{k'}\Gamma_{k'}\left\langle \hat{p}_{-k'}\hat{q}_{k'}-\xi_{\tilde{y},\tilde{z}}\hat{q}_{k'}\hat{p}_{-k'}\right\rangle \\
& +iM_{\tilde{x},\tilde{z}}\frac{1}{\sin \theta}\frac{1}{2}\kappa\left(1-\xi_{\tilde{x},\tilde{z}}\right)\left(1-\varepsilon+\delta_{\alpha}^{qq}-\frac{1}{Ns}n_{k=0}+\frac{1}{Ns}\Delta_{k=0}^{qq}\right)
\end{align}
\begin{align}
dQ_{\mathcal{D}_{D}}  & =M_{\tilde{y},\tilde{y}}\kappa\frac{\Gamma_{k}}{\Gamma_{k=0}}+iM_{\tilde{x},\tilde{y}}\kappa\left(1-\xi_{\tilde{x},\tilde{y}}\right)\frac{\Gamma_{k}}{\Gamma_{k=0}}\Delta_{k}^{qq}
\end{align}
\begin{align}
dP_{\mathcal{D}_{D}} & =M_{\tilde{x},\tilde{x}}\kappa\frac{\Gamma_{k}}{\Gamma_{k=0}}+iM_{\tilde{x},\tilde{y}}\kappa\left(1-\xi_{\tilde{x},\tilde{y}}\right)\frac{\Gamma_{k}}{\Gamma_{k=0}}\Delta_{k}^{pp}
\end{align}
\begin{align}
dW_{\mathcal{D}_{D}}  & =iM_{\tilde{x},\tilde{y}}\kappa\frac{\Gamma_{k}}{\Gamma_{k=0}}\frac{1}{2}\left\langle \hat{q}_{k}\hat{p}_{-k}-\xi_{\tilde{x},\tilde{y}}\hat{p}_{k}\hat{q}_{-k}+\hat{q}_{-k}\hat{p}_{k}-\xi_{\tilde{x},\tilde{y}}\hat{p}_{-k}\hat{q}_{k}\right\rangle 
\end{align}
\end{widetext}
Note that the spin-wave density is expressed in terms of two-point correlation functions as $\varepsilon\left(t\right)=\frac{1}{Ns}\sum_{k\neq0}n_{k}$ where $n_{k}=\frac{1}{2}\left(\Delta_{k}^{qq}+\Delta_{k}^{pp}-1\right)$. After assembling the contributions of each desired channel to the equations of motion for the collective spin angles and two-point functions, we then plug in the final expression for $\frac{d}{dt}\phi$ into the Larmor term in the equations of motion for the two-point functions. We then keep terms that are second order in $k\neq0$ spin-wave operators. As each Larmor term is proportional to $\frac{d}{dt}\phi$ multiplied by a two-point function, we only keep terms in $\frac{d}{dt}\phi$ that are zeroth order in spin-wave operators when substituting the expression. In the above expressions, we have kept terms that are proportional to $\frac{1}{Ns}$ which are necessary to quantify finite size effects. In the thermodynamic limit, these terms vanish. The treatment thus results in a set of differential equations for the collective angles $\theta(t)$ and $\phi(t)$ which are coupled to the $2N$ equations of motion for the two-point correlation functions which represent the dynamics of spin-wave excitations. The coupling between these equations represents the self-consistent part of the method where the quantum fluctuations of spin-waves affects the motion of the spin-wave vacuum and vice-versa.

The initial values of the dynamical variables depend on the choice of initial state. For quasiclassical pure states fully polarized along a given direction, the dynamical variables take the values $\theta(0)=\theta_0$, $\phi(0)=\phi_0$, $\Delta_k^{qq}=\Delta_k^{pp}=1/2$, and  $\Delta_k^{qp}=0$.

\section{Model}
\label{sec:model}


We now introduce a specific spin model which exhibits novel correlation dynamics illustrative of spatially extended dissipation. The system is described via the following purely dissipative non-diagonal Lindblad quantum master equation:
\begin{equation}\label{mainlind}
\partial_t\hat{\rho}=K\sum^N_{n,m=1 }f_{n,m}\left(\hat{S}_{n}^{-}\rho\hat{S}_{m}^{+}-\frac{1}{2}\{\hat{S}_{n}^{+}\hat{S}_{m}^{-},\rho\}\right).
\end{equation}

This model, with a tunable profile $f_{n,m}$, can be experimentally realized in ensembles of two-level atoms coupled to a cavity mode as described in Ref.~\cite{SeetharamShort}, where it is also shown that the correlations generated by this dissipation can be modified into novel spatio-temporal patterns by a coherent uniform external field acting on the system.
%
%
The spatial extension of the dissipation is contained in the translationally-invariant  profile $f_{n,m}=f(\abs{n-m})$, while its strength $K\equiv  {2\kappa}/{( \Gamma_{k=0})}$ 
is renormalized by $\Gamma_{k=0}$ where $\Gamma_{k}\equiv\sum_{r\in\left\{ -\frac{N}{2},\frac{N}{2}\right\} }e^{ikr}f\left(\left|n-m\right|\right)$ is the Fourier transform of $f_{n,m}$. 

In the language of Sec.~\ref{sec:TDSW}, the system described by Eq.~\eqref{mainlind} has observables $\hat{A}$ that evolve according to the adjoint master equation $\frac{d}{dt}\hat{A}=\mathcal{D}_{D}\left(\hat{A}\right)$ with:
\begin{widetext}
	\begin{equation}\label{eq:MainDispChannel}
	\mathcal{D}_{D}\left(\hat{A}\right)   =\frac{\kappa}{s\Gamma_{k=0}}\sum_{n,m}f\left(\left|n-m\right|\right)\left(\hat{S}_{n}^{+}\hat{A}\hat{S}_{m}^{-}-\frac{1}{2}\left\{ \hat{S}_{m}^{+}\hat{S}_{n}^{-},\hat{A}\right\} \right)
	\end{equation}
\end{widetext}

Note that we drop the superscript $D$ in this section. It is instructive to analyze dynamics described by Eq.~\eqref{eq:MainDispChannel} starting from the case of a long-range spatial profile, $f\left(\left|n-m\right|\right)=(\abs{n-m}+1)^{-\alpha}$. The Fourier transform, $\Gamma_{k}$, of this profile can be expressed in terms of poly-logarithm functions  $\Gamma_{k}(\alpha)=1+2\text{Re}\left[e^{-ik} \text{Li}_{\alpha}\left(e^{ik}\right)-1\right]$  of order $\alpha$.
The inclusion of a hardcore parameter $R=1$ ensures positivity of $\Gamma_k$, as explained in Sec.~\ref{sec_types} and further in Appendix~\ref{sec:appdx_positivity}
The denominator $\Gamma_{k=0}$ ensures the extensive scaling of the Liouvillian~\eqref{mainlind} in the thermodynamic limit, thus playing a role analogous to the conventional Kac's renormalization of long-range Hamiltonians~\cite{Ruffo,kastner,Daley,vodola2014kitaev, vodola2015long}.
%

When $\alpha=0$, the dynamics of the collective spin admit an analytical solution in the thermodynamic limit~\cite{Fazio,prazeres2021boundary}. The mean-field solution becomes exact and can be written in terms of the components of the collective magnetization, $m^x(t)=\sin\theta(t)\cos\phi(t)$ and $m^z(t)=\cos\theta(t)$, which, in this case, is fully described by a spin coherent state moving on the (collective spin) Bloch sphere with azimuthal and polar angles $\phi(t)$ and $\theta(t)$, respectively. The model at $\alpha=0$, with the addition of a coherent external field representing by a Hamiltonian $\hat{H}_{0}=\omega_0\sum_{n=1}^{N}\hat{S}_{n}^{x}$, has been studied in the context of cooperative radiation, optical bistability, and time-crystals~\cite{Fazio,PhysRevA.22.1179, drummond1978volterra, walls1978non, hannukainen2018dissipation}. When $\omega_0/\kappa\gtrsim 1$, the total magnetization rolls around the $\hat{x}$ axis with $\langle \hat{S}^z\rangle=0$. In the opposite limit  $\kappa/\omega_0\gtrsim 1$, the dynamics is damped and quickly attracted towards the southern hemisphere of the Bloch sphere with a non-vanishing $\hat{S}^z$ component. 

Choosing $\alpha\neq0$ introduces spatial resolution to the system and understanding the dynamics requires, in principle, a solution to the full many-body system including connected spin correlation functions of all orders beyond mean-field.
In the dissipation-dominated regime $\kappa/\omega_0\gtrsim 1$, however, the NEQSWT developed in Sec.~\ref{sec:TDSW} can be used to treat the system as the number of spin-wave excitations remains sufficiently low over the course of dynamics. In the next section, we analyze dynamics for a system with no external field ($\omega_0=0$). As the dissipation channel, Eq.~\eqref{mainlind}, is the only generator of dynamics, we are always in the dissipation-dominated regime where NEQSWT remains valid. The case of a non-zero external field ($\omega_0\neq0$) is discussed in Ref.~\cite{SeetharamShort}, with the overall picture unaffected by a small but non-zero $\omega_0$.

For $0\le \alpha\le 1$, the normalization factor diverges, which means that the normalized spectrum of the Lindbladian $(\kappa/s)\Gamma_k/\Gamma_{k=0}$ asymptotically converges to zero for all finite momenta $k\neq0$. 
It can be shown that such a spectrum remains discrete in the thermodynamic limit~\cite{lerose2020origin,defenupnas}.
Away from fine-tuned dynamical critical points, however, the behavior of collective observables in the thermodynamic limit is
identical with that of the mean-field model describing the $\alpha=0$ case. There may be severe finite-size effects for $\alpha$ close to $1$.
For $\alpha>1$, however, spin-wave modes get populated as the system evolves out of equilibrium, exerting a finite feedback on the dynamics of the collective spin, which acquire corrections beyond the mean-field. In the next section, we consider this situation.

\section{Dynamics of correlation functions for short- and long-range losses}\label{sec:DispDynamics}


\begin{figure*}[t!]
	\centering
	\includegraphics[width=0.90\textwidth]{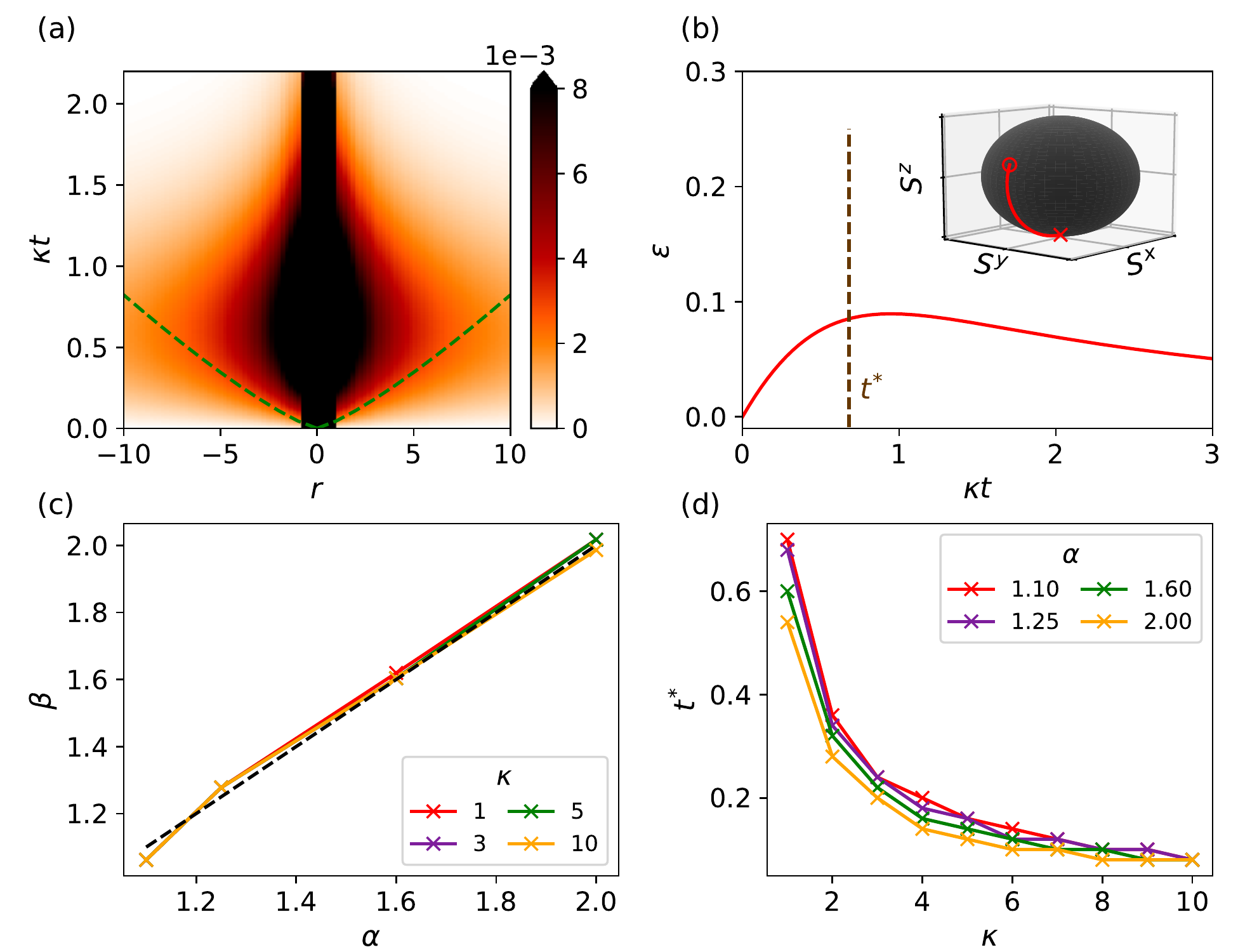}
	\caption{\textbf{Dynamics of $\hat{L}_{n}=\hat{S}_{n}^{-}$ dissipation with long-range spatial profile $f\left(\left|r\right|\right)=(\abs{r}+1)^{-\alpha}$.} \textbf{(a)} Spreading and contraction of spin correlations described by Eq.~\eqref{eq_czz} for $\alpha=1.25$ and  $\kappa=1.0$; the green dotted line tracks the correlation front which spreads as $t\approx r^\beta$ at short times.   \textbf{(b)} Dynamics of the spin wave density and evolution of the collective magnetization on the Bloch sphere (inset) for the same choice of parameters as (a). The density of spin waves has a peak at  time $t^*$ where the front of correlations   reverses (cf. (a)). \textbf{(c)} Scaling parameter $\beta$ as a function of $\alpha$. The black dotted line represents $\beta=\alpha$; we see that $\beta\simeq\alpha$  independent of the dissipation strength $\kappa$. \textbf{(d)} Dependence of $t^*$~on $\alpha$ and $\kappa$. For all panels we evaluate dynamics in the thermodynamic limit with the initial state of the system representing a spin coherent state pointing in the direction $\theta(t=0)=0.4\pi$, $\phi(t=0)=0$.}
	\label{fig:LRDispAnalysis}
\end{figure*}

\begin{figure*}[t!]
	\centering
	\includegraphics[width=0.90\textwidth]{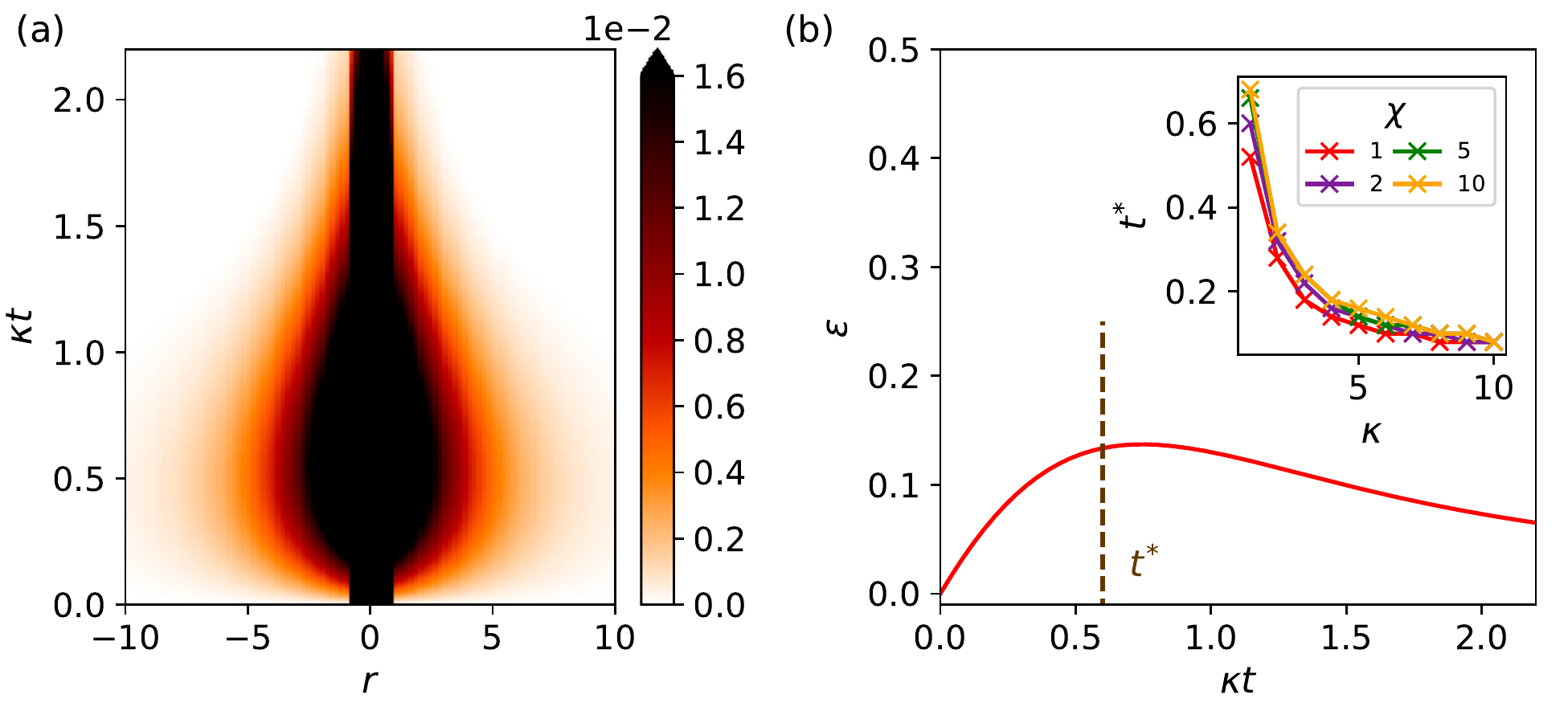}
	\caption{\textbf{Dynamics of $\hat{L}_{n}=\hat{S}_{n}^{-}$ dissipation with short-range spatial profile $f\left(\left|r\right|\right)=\exp(-\abs{r}/\chi)$.} \textbf{(a)} Spreading and contraction of spin correlations described by Eq.~\eqref{eq_czz} for $\chi=2.0$ and  $\kappa=1.0$. \textbf{(b)} Dynamics of spin-wave density and correlation function transition time (inset). For all panels we evaluate dynamics in the thermodynamic limit with the initial state of the system representing a spin coherent state pointing in the direction $\theta(t=0)=0.4\pi$, $\phi(t=0)=0$.}
	\label{fig:SRDispAnalysis}
\end{figure*}


We can gain intuition for the dynamics of quantum correlations generated by Eq.~\eqref{eq:MainDispChannel} by considering the simplest situation of evolution starting from a pure state with a  single spin excitation $\rho(0)= |j\rangle\langle j|$, where $|j\rangle \equiv |\downarrow_1\dots\downarrow_{j-1}\uparrow_j\downarrow_{j+1}\dots\downarrow_N\rangle$ and the single up spin is at site $j$. Time evolution takes place in the restricted Hilbert space spanned by $\{ |n\rangle, n=1,\dots,N \}$ and the dark state $|\emptyset \rangle\equiv |\downarrow_1\dots\downarrow_N\rangle$ which is fully polarized down. 
    The state's dynamics, governed by the master equation Eq.~\eqref{eq:MainDispChannel}, admits a simple physical picture: the excitation initially at site $j$ evolves in the single-excitation space subject to the non-Hermitian (imaginary) hopping Hamiltonian \begin{equation}
    \label{eq_singleexc}
  \frac{i\kappa}{2s\Gamma_{k=0}}\sum_{n,m} f_{|n-m|} L^\dagger_n L_m \; \leadsto \; \frac{i\kappa}{2s\Gamma_{k=0}}
        \sum_{n,m} f_{|n-m|} |n\rangle\langle m| \,.
    \end{equation} 
    This single-particle evolution is diagonal in Fourier space, and each momentum component decays with a different rate proportional to $\Gamma_k/\Gamma_{k=0}$.
    The probability continuously lost by the single-excitation space accumulates in the dark state.
    Thus, the excitation initially localized at site $j$ spreads quantum mechanically in the single-excitation sector, generating an initial  spreading of quantum correlations with quantum coherence between different single-excitation states. At the same time, each momentum component inhomogeneously decays to the dark state, which generates nontrivial correlation dynamics~\cite{SeetharamShort}.

Armed with intuition from the above example, we proceed to examining the dynamics of Eq.~\eqref{mainlind} starting from initial states far away from the dark state. We will consider systems prepared in fully polarized states pointing along an arbitrary direction on the Bloch sphere, identified by spherical angles $\theta_0$, $\phi_0$. Furthermore,
we consider long-range and short-range spatial profiles $f(r=\abs{n-m})$ given respectively by 
\begin{equation}
f\left(r\right)=\frac 1 {(r+1)^{\alpha}} \qquad \textrm{or} \qquad
f\left(r\right)=\exp(-r/\chi). 
\end{equation}
%
Using Eqs~\eqref{eq:gaussiane}, we can derive a differential equation for the occupation $n_k$ of the spin-wave excitation at wavevector $k\neq0$.
\begin{equation}\label{eq:SWEoM}
\frac{d}{dt}n_{k}=2\kappa\frac{\Gamma_{k}}{\Gamma_{k=0}}\left(n_{k}\cos\theta(t)+\cos^{4}\left(\frac{\theta(t)}{2}\right)\right).
\end{equation}
The $k$-dependent prefactor $\Gamma_{k}/\Gamma_{k=0}$ is positive for both spatial profiles of interest; 
positivity of the master equation requires $\kappa\ge 0$. Remarkably, for the specific Lindblad channel in Eq.~\eqref{eq:MainDispChannel}, the equation of motion for spin-wave occupation given by Eq.~\eqref{eq:SWEoM}  is a linear differential equation that is not coupled to other NEQSWT variables. The homogeneous term in Eq.~\eqref{eq:SWEoM} describes the rate of production of spin-waves and depends on $\cos\theta(t)$; accordingly, it generates or drains spin waves depending on whether the collective magnetization is in the northern ($0<\theta(t)<\pi/2$) or southern ($\pi/2<\theta(t)<\pi$) hemisphere of the Bloch sphere. In other words, the transition in the rate of production of spin waves can be understood as a consequence of the spin waves' dynamics being dependent on the instantaneous direction of the collective spin. While the effect of dissipation is creating spin waves on top of a mean field in the northern hemisphere, the same dissipative mechanism results in a reduction of spin-waves with respect to a mean-field in the southern hemisphere.
We note that the inhomogeneous decay of spin excitations with different momenta, discussed at the beginning of this section [cf. Eq.~\eqref{eq_singleexc}], is visible in Eq.~\eqref{eq:SWEoM} by examining the point $\theta\simeq \pi$.

Note that this behavior is a result of the choice of dissipation channel, $\hat{L}_{n}=\hat{S}_{n}^{-}$, and does not depend on the choice of spatial profile which only modifies the prefactor $\Gamma_k/\Gamma_{k=0}$ in Eq~\eqref{eq:SWEoM}. 
The long-range profile is a power-law decay characterized by power $\alpha$ and results in a prefactor that decays as a a power-law with power related to $\alpha$. The short-range profile is an exponential decay characterized by a decay length $\chi$ and results in a prefactor that is Lorentzian with width proportional to $1/\chi$. The change in spatial profile determines modifications in some non-universal parameters such as the transition time, $t^*$ upon which the system switches from pumping excitations to draining excitations. The spatial profile is, however, important when engineering the dynamics of the system for certain applications~\cite{SeetharamShort}.

The mechanism governing the dynamics of spin-wave occupation explains  the dynamics of equal time spin-spin correlation functions. As an example, we examine the connected correlation function 
\begin{equation}
\label{eq_czz}
	C^{zz}\left(r,t\right)=\langle \hat{S}_{n}^{z}\left(t\right)\hat{S}_{n+r}^{z}\left(t\right)\rangle -\langle \hat{S}_{n}^{z}\left(t\right)\rangle \langle \hat{S}_{n+r}^{z}\left(t\right)\rangle
\end{equation}
which is directly sensitive to the action of spin losses $\hat{L}_n={\hat{S}_n}^-$. This function can be expressed in terms of NEQSWT variables as
\begin{equation}\label{eq:CzzTDSW}
	C^{zz}\left(r,t\right) =\left(\sin\theta(t)\right)^{2}\sum_{k\neq0,k>0}\cos(kr)\Delta_{k}^{qq}.
\end{equation}
We see that there is an overall envelope to the correlation dynamics set by $\left[\sin\theta(t)\right]^{2}$, which grows as the collective spin moves from the north pole of the Bloch sphere to the equator, and shrinks as it moves from the equator to the south pole. Therefore, in the absence of other dynamical channels, we expect the correlations to grow for a period of time and then shrink, with the time $t^{*}$ upon which the system transitions between these two regimes being dependent on the motion of the collective spin. As the dynamics of spin-wave occupation also increases and decreases depending on the collective spin motion, we expect that the correlation transition time $t^{*}$ sets the scale upon which the spin-wave density $\varepsilon$ reaches its maximum value before shrinking. Similar to the dynamics of spin-wave occupation, we note that the choice of spatial profile does not qualitatively modify the correlation dynamics. The spatial profile only enters Eq.~\eqref{eq:CzzTDSW} through the dynamics of $\Delta_{k}^{qq}$.

We now numerically calculate the dynamics of the correlation function, Eq.~\eqref{eq:CzzTDSW}, using NEQSWT and analyze both long-range and short-range cases. We start with all the spins in a coherent state pointing slightly above the equator of the Bloch sphere ($\theta(t=0)=0.4\pi$, $\phi(t=0)=0$). The qualitative nature of the dynamics for this dissipative channel does not depend on the angle of the initial coherent state; starting too close to the North pole, however, causes the spin-wave density to exceed the threshold treatable by NEQSWT. Our choice of $\theta(t=0)=0.4\pi$ allows the dynamics to be validly treated with NEQSWT. 

The correlation dynamics for the long-range spatial profile is shown in Fig.~\ref{fig:LRDispAnalysis}(a). In the first stage of dynamics, correlations exhibit a front scaling as  $t\approx r^\beta$. The exponent $\beta$ is plotted in Fig.~\ref{fig:LRDispAnalysis}(c), showing that the dissipation strength $\kappa$ does not play a role in the `opening' of the correlation function. The exponent $\beta$ characterizing the scaling follows $\beta\simeq\alpha$; this result can be understood by making the following scaling ansatz for $C^{zz}\left(r,t\right)$ in the initial opening stage of correlation spreading dynamics:
\begin{equation}\label{scale0}
C^{zz}\left(rt_{1}^{1/\beta},t_{1}\right)=C^{zz}\left(rt_{2}^{1/\beta},t_{2}\right).
\end{equation}
Algebraic manipulation yields the equivalent expressions
\begin{equation}\label{scale}\begin{split}
C^{zz}\left(\zeta r,t\right)&=\zeta^{\nu} C^{zz}\left(r,t\right),\\
C^{zz}\left(r,\zeta t\right)&=\zeta^{-\nu \eta} C^{zz}\left(r,t\right).
\end{split}\end{equation}
Here $\zeta$ is a positive rescaling factor while $\nu$ and $\eta$  are the two rescaling exponents for space and time. The above ansatz represents a correlation function front scaling with exponent $\beta=1/\eta$.
As we discuss later, we find that for large distances ($r\gg1$), the correlation function satisfies $C^{zz}\left(r,t\right)\propto1/r^\alpha$. This behavior yields $\nu=-\alpha$ using the first equation in~\eqref{scale}. Additionally, at short times, correlations grow linearly to leading order ($C^{zz}\left(r,t\to0\right)\propto t + \mathcal{O}\left(t^{2}\right)$) as we start with an uncorrelated spin coherent state for which $C^{zz}\left(r,t=0\right)$ is vanishing. The second equation in~\eqref{scale} therefore implies $\nu \eta=-1$ and combining them,  yields $\eta=1/\alpha$. We therefore see that the correlation front must scale as $t\simeq r^\beta$ with $\beta=\alpha$ as numerically observed.
At large $\alpha$, correlations disappear ($\beta\to\infty$)  consistently with the Lindbladian becoming diagonal and representing independent local emission events. This behavior differs from the large $\alpha$ light cone of long-range Hamiltonians which becomes increasingly linear ($\beta\approx1$)~\cite{PhysRevLett.119.190402}. As stated in Sec.~\ref{sec:TDSW}, this difference arises from the proper way to define  long-range  dissipation ($f\left(\left|n-m\right|\right)=(\left|n-m|+1\right)^{-\alpha}$) versus coherent dynamics ($f\left(\left|n-m\right|\right)=\left|n-m\right|^{-\alpha}$). In the former case, we tend towards independent dissipators for large $\alpha$, while in the latter case one retrieves  nearest-neighbor interactions. Similar phenomenology is retrieved for short-range losses when $\chi\to0$.

At late times, long-range dissipation has a contractive effect on correlation dynamics. Correlations reach their maximum spread at a time $t^*$ where the spin wave density exhibits  a peak. Spin waves are pumped by the second term in the right hand side of Eq.~\eqref{eq:SWEoM} which acts as parametric drive, and they are damped by the first term of~\eqref{eq:SWEoM} as soon as the  collective magnetization enters the southern hemisphere. For sufficiently strong dissipation, the collective magnetization will always eventually enter the southern hemisphere as the south pole is the dark state for strong spin losses. The competition of this self-pumping mechanism and the incoherent depolarization of spins is what leads to the opening and closing of the correlation function. The transition time $t^*$ corresponds to the timescale upon which the spin wave damping term starts to dominate dynamics (see Fig.~\ref{fig:LRDispAnalysis}(d)). Correlations vanish in the absence of spin wave excitations and therefore the correlation function $C^{zz}\left(r,t\right)$ shrinks to zero as spin waves are progressively dissipated into the environment for $t> t^*$ (see Fig.~\ref{fig:LRDispAnalysis}(b)). At sufficiently late times ($t\gg t^*$), there is negligible spin wave density and the system is almost in a coherent state of spins pointing in a direction near the south pole. Closer inspection into the correlations near the steady state shows that $ C^{zz}(r)\propto 1/r^\alpha$ for large inter-spin distances. In fact, this $1/r^\alpha$ decay of correlations appears to hold at all times.

We also examine the correlation dynamics for a short-range spatial profile. Figure~\ref{fig:SRDispAnalysis}(a) shows that the correlations follow the same qualitative behavior as the the long-range case (they grow for a period before contracting). The time $t^{*}$ characterizing this transition is shown in Fig.~\ref{fig:SRDispAnalysis}(b) and it corresponds to the time upon which spin-wave excitations reach their maximal value and start decreasing. In both long- and short-range cases, the time scale $t^{*}$ increases for   spatial profiles that decay more slowly in space. However, the dependence on spatial profile is weak and the transition time   primarily depends on the decay rate $\kappa$ which sets the overall time-scale of the dissipation channel. The main difference between long- and short-range dissipative dynamics is that the correlations decay more rapidly in space for the short-range case, as seen by comparing Fig.~\ref{fig:LRDispAnalysis}(a) to Fig.~\ref{fig:SRDispAnalysis}(a).

\section{Future directions  }
\label{fut}

In this work, we have characterized the spatio-temporal spread of correlations generated by  dissipation with both short- and long-range spatial profiles, focusing on systems initialized in uncorrelated coherent spin states. Comparing how correlations spread when generated by spatial extended dissipation versus coherent interactions may enable discovery of novel classes of quantum information transfer phenomena. 

Our analysis was made possible by generalizing the formalism of NEQSWT. There are several interesting directions that could be explored with further methodological improvements. For example, we plan to extend the generalized NEQSWT to a Hartree-Fock treatment of non-linear effects beyond the leading order Holstein-Primakoff expansion. This would allow us to analyze systems with sizeable spin-wave densities, enabling the study of systems with highly correlated initial states, as well as exploring the possibility of dynamical phase transitions arising from competition between unitary dynamics generated by a Hamiltonian and dissipative dynamics generated by a Lindblad channel.

An experimental implementation of the model studied in this work, Eq.~\eqref{mainlind}, was proposed in a cavity QED platform of atoms trapped in a very leaky cavity~\cite{SeetharamShort}. In order to provide a closer benchmark with cavity QED experiments and explore regimes where coherent and dissipative dynamics of the cavity compete, a method to treat the combined light-matter system is required. We envision the possibility of extending variational many-body methods~\cite{shi2018variational} to study how correlations spread in the system when the cavity photon cannot be adiabatically eliminated and will therefore participate in the dynamics of the atoms. When the photon linewidth is decreased, the spatio-temporal spin correlation patterns may get modified in non-trivial ways~\cite{kelly2021effect}.

%

%

\begin{acknowledgments} 
 KS conducted this research with Government support under and awarded by DoD, Air Force Office of Scientific Research, National Defense Science and Engineering Graduate (NDSEG) Fellowship, 32 CFR 168a, and NSF EAGER-QAC-QCH award No. 2037687.
{AL acknowledges support from the Swiss National Science Foundation.}
RF acknowledges partial financial support from the Google Quantum Research Award. R.F.'s work has been conducted within the framework of the Trieste Institute for Theoretical Quantum Technologies (TQT). JM was supported by the European Union's Framework Programme for Research and Innovation Horizon 2020 under the Marie Sklodowska-Curie Grant Agreement No. 745608~(`QUAKE4PRELIMAT'). 
JM acknowledges support from  the Deutsche Forschungsgemeinschaft (DFG, German Research Foundation)   Project-ID 429529648 TRR 306 QuCoLiMa (Quantum Cooperativity of Light and Matter).
\end{acknowledgments}

\appendix

\section{Lindblad positivity constraint}\label{sec:appdx_positivity}

\renewcommand{\thefigure}{A\arabic{figure}}
\setcounter{figure}{0}

\begin{figure}[h!]
	\centering
	\includegraphics[width=0.99\columnwidth]{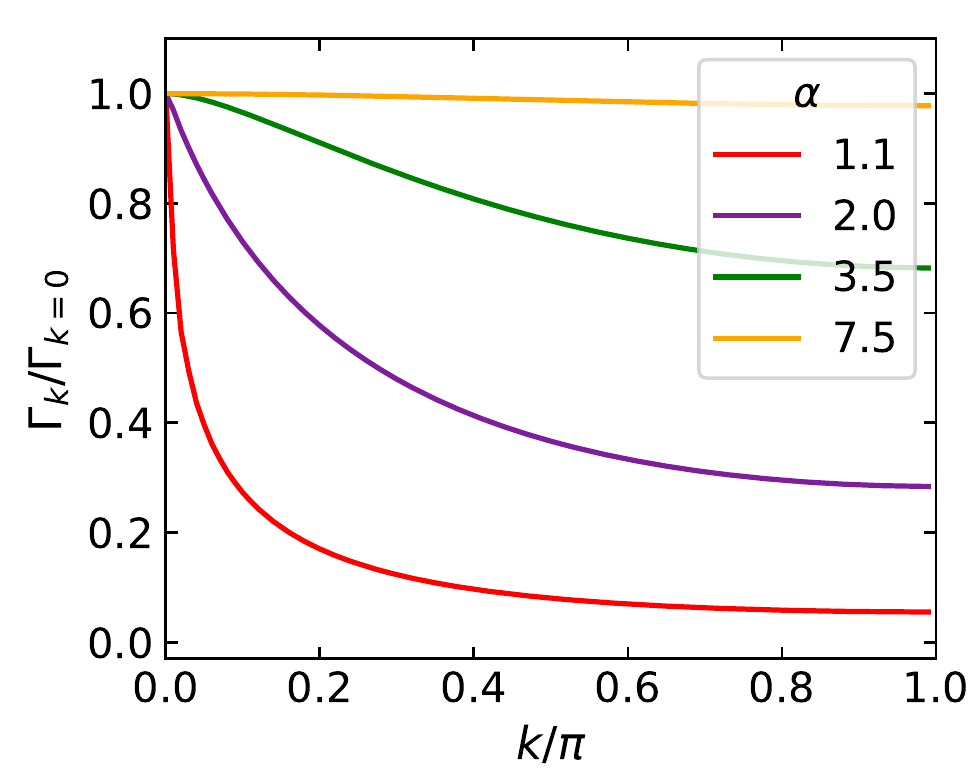}
	\caption{\textbf{Fourier transform of long-range spatial profile $f\left(\left|r\right|\right)=(\abs{r}+1)^{-\alpha}$.} The function is symmetric across $k=0$ for $k\in\left[-\pi,0\right]$. The fact that $\Gamma_k$ is greater than zero for all $\alpha>1$ ensures that a Lindblad channel with this spatial profile is mathematically well-defined.}
	\label{fig:FigAppdx}
\end{figure}

The spatial profile $f\left(\left|n-m\right|\right)$ of a valid Lindblad map must be positive semi-definite, which translates to the requirement that the Fourier transform of the function $f(r) := f_{r=|n-m|}$ is a non-negative function, i.e., 
$$\widetilde{f}(k) = \sum_{r=-\infty}^{+\infty} e^{-ikr} f(r) \ge 0 \, .$$
This is because translational invariance implies that $\{ \widetilde{f}(k) , -\pi < k \le \pi \}$ are proportional to the eigenvalues of dissipator. Spatial profiles of dissipative Lindblad channels must thus be defined such that they satisfy this constraint. For example, a long-range spatial profile can be properly defined for a dissipative channel as $f(r)=1/(1+|r|)^\alpha$. We show in Fig.~\ref{fig:FigAppdx} that the Fourier transform of this profile is positive for all $\alpha>1$, which is the regime where spatial correlations survive in the thermodynamic limit.

The positivity condition can also be proved analytically.
Consider the more general form of a long range spatial profile
\begin{equation}
    f(r)=\frac 1 {(R+|r|)^\alpha}
\end{equation}
with a tunable hardcore parameter $R$. Let us separate the effects of the local part $f^L(r)= \delta_{r,0} \frac 1 {R^\alpha}$ and the non-local part $f^{NL}(r)= (1-\delta_{r,0}) \frac 1 {(R+|r|)^\alpha}$ of the dissipation spatial profile.
The Fourier transform is
\begin{equation}
\widetilde{f}(k) =
\widetilde{f}^L(k)
+
\widetilde{f}^{NL}(k)
=
f(0) + 
2 \sum_{r=1}^{+\infty} \cos(kr) f(r) \, .
\end{equation}
Note that by construction
\begin{equation}
\int_{-\pi}^\pi \frac {dk}{2\pi} \widetilde{f}^{NL}(k) = {f}^{NL}(0) = 0
\end{equation}
so the nonlocal part $\widetilde{f}^{NL}(k)$ is equally distributed above and below zero.
The local part $\widetilde{f}^{L}(k)$ is a positive additive constant equal to $f(0)$.
Thus, we can choose the value of $f(0)$ to push the full Fourier transform entirely up above the horizontal axis, thereby realizing positivity. 
The smaller the $R$, the larger this constant.
A simple sufficient criterion can be proven as follows: \\
one can choose $R=R(\alpha)$ such that the last inequality holds:
\begin{equation}
f(0)+\min_k \widetilde{f}^{NL}(k) \ge
f(0)-\max_k |\widetilde{f}^{NL}(k)| \ge 0 \, .
\end{equation}
To do so, we bound
\begin{multline}
\max_k |\widetilde{f}^{NL}(k)| \le 2 \sum_{r=1}^\infty \frac{1}{(R+r)^\alpha} \\
\le 2 \int_0^\infty \frac{dx}{(R+x)^\alpha}
=
\left( \frac {2R} {\alpha-1} \right) \frac 1 {R^\alpha}.
\end{multline}
Positivity is guaranteed when this quantity does not exceed $f(0)=1/R^\alpha$. Thus, we obtain the sufficient criterion
\begin{equation}
R \le \frac {\alpha-1} 2
\, .
\end{equation}
This bound is not tight, because we majorized $|\min_k \widetilde{f}^{NL}(k)| $ 
by 
$\max_k |\widetilde{f}^{NL}(k)|$: For all $1\le\alpha<\infty$, the former extremum is realized at $k=\pi$ and the latter at $k=0$, so the two quantities are always different.
In reality, the value $R=1$ that we chose in our study is sufficient for all $\alpha$'s: for $R=1$ one has
\begin{equation}
\widetilde{f}(k) = 1
+ \text{Re}
\bigg[
\sum_{r=1}^\infty \frac{e^{ikr}}{(1+r)^\alpha}
\bigg]
=
1 + \text{Re}
\Big[
e^{-ik} \text{Li}_\alpha(e^{ik}) - 1 \Big] \, ,
\end{equation}
where $\text{Li}_\alpha(z)=\sum_{r=1}^\infty z^r/r^\alpha$ is the polylogarithmic function.
The function on the right-hand side is positive in the whole domain $k\in(-\pi,\pi]$, $\alpha\in[1,\infty)$.


%

\end{document}